# Molecular Communication: Physically Realistic Models and Achievable Information Rates

## Andrew W. Eckford

### Abstract

Molecular communication is a biologically-inspired method of communication with attractive properties for microscale and nanoscale devices. In molecular communication, messages are transmitted by releasing a pattern of molecules at a transmitter, which propagate through a fluid medium towards a receiver. In this paper, molecular communication is formulated as a mathematical communication problem in an information-theoretic context. Physically realistic models are obtained, with sufficient abstraction to allow manipulation by communication and information theorists. Although mutual information in these channels is intractable, we give sequences of upper and lower bounds on the mutual information which trade off complexity and performance, and present results to illustrate the feasibility of these bounds in estimating the true mutual information.

## I. INTRODUCTION

At the scale of microorganisms, which have dimensions on the order of $10^{-6}$ meters, biological communication is often carried out using molecules; for example, it is well known that chemical signals are used in communication and control among cells in living tissue. As another example, some species of bacteria send chemical messages to their neighbors in a strategy known as "quorum sensing", which allows them to estimate the local population of their own species [1]. Given these examples, *molecular communication* [2] has been proposed as a biologically inspired solution to the problem of communicating among microscale or nanoscale devices. Molecular communication is an engineered form of communication, using designed systems to send messages from a transmitter to a receiver: the transmitter sends a message by releasing a pattern of molecules into a shared medium, while the receiver detects the the message by observing the arriving pattern of molecules, similarly to the related biological systems. The objective of this paper is to provide an analytical basis for molecular communication by providing physically realistic models and methods for bounding on information rate.

Molecular communication has attracted considerable attention from researchers across several disciplines, including electrical engineering, microbiology, chemistry, and biomedicine. However, unlike most conventional forms of communication, virtually all of the molecular communication results in the literature are experimental rather than analytical. There are many examples of systems that have been implemented in laboratory settings: for example, recent successes in the new field of *systems biology* [3], in which microorganisms are engineered and designed to perform specific tasks, have been exploited to produce molecular communication systems. Fundamental work in this direction was done by Weiss [4], [5], who adapted chemical pathways in microorganisms to construct simple "circuits" that communicate with each other, such as logic gates. Other researchers have focused on biological components that can be exploited in molecular communication. For example, to communicate between two engineered cells that are in direct contact, *gap junctions* (i.e., portals through two adjacent cell membranes) may be used to pass message-bearing molecules; a communication system using this principle was described in [6], [7]. Another technique packages message-bearing molecules in a small container known as a *vesicle*, and conveys them along a filament connecting two devices using a *molecular motor* [8], [9]. Since these two techniques require a connection between communicating devices, either direct contact or via a filament, they are analogous to wired communication. On the other hand, molecules may propagate in free space between the transmitter and receiver via Brownian motion, such as in the system proposed by [10]; this requires no connection and is analogous to wireless communication. Extensive experimental work has been conducted into molecular communication and related methods, and the references listed above are only a representative sample of that work.

Meanwhile, very little work has been produced to provide an information-theoretic or communication-theoretic analysis of these channels. Berger [11] has presented a related idea known as "living information theory", where the goal was to analyze biological systems using information-theoretic tools. Work has also been done to find the capacity of the so-called *chemical channel*, also known as the *trapdoor channel* [12], [13]. This model captures many interesting features of the molecular communication problem, but it is inadequate: for instance, as we show in Section IV-B, for any finite number of balls in the bin, there exist sequences of molecule arrivals that occur with $\Pr > 0$ in practice, but that have $\Pr = 0$ using the trapdoor channel. In terms of related work, similarities exist with timing channels, such as the queue channel [14], [15].

Our main results are concerned with mathematical modeling of molecular communication, and information-theoretic performance bounds; these results are summarized as follows:

1) **Modeling.** We present physically realistic models for the propagation environment, as well as an *ideal model* of the transmitter and receiver in any molecular communication system (Section II). These models provide a level of abstraction such that they may be used by information and communication theorists, who may have no particular background in chemistry or biology, to analyze and design molecular communication systems. We show that our ideal model is information-theoretically meaningful, in that it provides an upper bound in terms of mutual information for any alternative model (Theorem 1). Further, we show that any system with distinguishable molecules can be separated into statistically independent systems (Theorem 2), so that from an information-theoretic perspective, it is sufficient to solve the case where molecules are indistinguishable.

2) **Performance bounds.** We give methods to find bounds on achievable information rate in molecular communication systems. When the input distributions are unconstrained, we show that the achievable rates are infinite (Theorem 3). For constrained input distributions, the mutual information is intractable, so we provide a sequence of both tractable and achievable lower bounds (Theorem 4, reused from [16] and elsewhere), based on a straightforward and extensible approximation of the channel (Section IV-D). We also provide a sequence of tractable upper bounds (Theorem 5). Both sequences of bounds provide a natural tradeoff between performance and complexity. Results are obtained to illustrate both the lower and upper bounds, which are given in (Section VI).

Throughout this paper, we focus on molecular communication systems using free-space diffusion, but our models can be easily generalized to a wide variety of alternative scenarios. With these results, our hope is to generate interest in molecular communication from information theorists, and to inspire further research in this emerging field.

The remainder of the paper is organized as follows. In Section II, we outline our channel model and provide some useful modeling results. In Section III, we give some simple results showing that information rates are infinite unless the input distribution is appropriately constrained. In Section IV, we give some nontrivial and achievable lower bounds on information rate, while in Section V, we give upper bounds. Finally, in Section VI, we give results illustrating the usefulness of these bounds.

## II. MODEL

In this section, we present a formal and physically realistic mathematical model for molecular communication based on free-space diffusion and Brownian motion. This model has useful properties for information theoretic analysis: first, it provides a layer of abstraction such that researchers with no background in chemistry can analyze molecular communication; and second, as we show in Theorem 1, the model is information-theoretically ideal, in the sense that relaxing our modeling assumptions leads to a system with smaller mutual information.

### A. Brownian motion as a communication medium

*Brownian motion* refers to the random motion of particles, which may be individual molecules, as a result of random collisions and interactions with molecules in the environment. There exists an extensive body of literature concerning Brownian motion as a stochastic process, which has applications in physics and beyond; the reader is directed to [17] for an introduction.

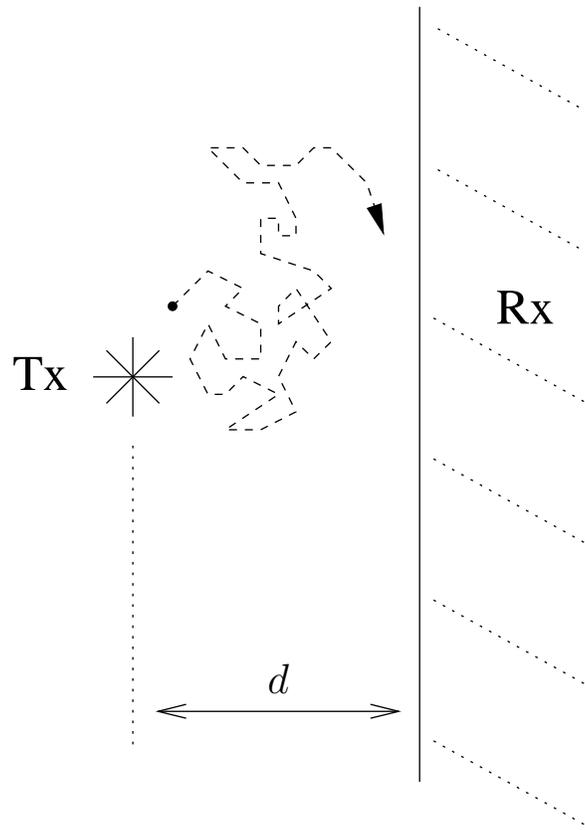

Fig. 1. The molecular communication system, with a point source transmitter, separated from the receiver by a distance $d$.

To see how Brownian motion may be used as a communication medium, suppose a point-source transmitter, at the origin, and a receiver, located distance $d$ away, are connected by a fluid medium, as depicted in Figure 1. For convenience, we assume a one-dimensional medium, but this is not essential to the remainder of the paper; further, as shown in the figure, the 1-d model is practical for a receiver that can be viewed as a plane (e.g., for a cell that is in close proximity to the transmitter). The transmitter has a message $w \in \mathcal{W}$, where $\mathcal{W}$ is the set of all possible messages. The transmitter conveys $w$ to the receiver by releasing a pattern of molecules into the fluid medium. The receiver observes the arrivals of the molecules, and from the pattern of arrivals, guesses that $w'$ was the message sent by the transmitter. As in any communication channel, if $w = w'$, the transmission is successful; if $w \neq w'$, an error is made.

For the remainder of the paper, we make three modeling assumptions about the Brownian motions and physical properties of the molecules: first, that they are *Markovian* (i.e., given the molecule's position and physical state in the present, its motion in the future is conditionally independent of its motion in the past); second, that molecules do not react or otherwise change in transit; and third, that they are statistically

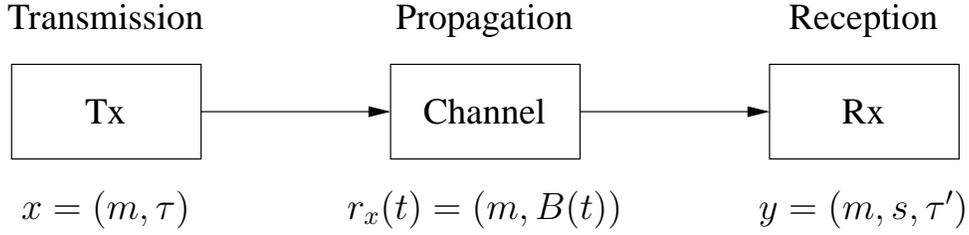

Fig. 2. Transmitter, channel, and receiver, with associated quantities.

independent for different molecules.

As depicted in Figure 2, a particular molecule undergoes three processes: *transmission*, *propagation*, and *reception*. Here we define these processes in full generality; our later assumptions and results will allow us to simplify some of these expressions.

Letting $\mathcal{M}$ represent the set of molecules that are available within the system, a *transmission* $x$ is defined as the pair

$$x = (m, \tau) \in \mathcal{M} \times \mathbb{R}, \tag{1}$$

representing the release of molecule of type $m \in \mathcal{M}$ at time $\tau \in \mathbb{R}$.

Letting $\mathcal{B}$ represent the set of possible Brownian motions $B(t)$, where $B(t)$ is the position of the Brownian motion as a function of time, the *propagation* of the transmission $x$ is defined as the double

$$r_x(t) = (m, B(t)) \in \mathcal{M} \times \mathcal{B} \tag{2}$$

representing the motion $B(t) \in \mathcal{B}$ of a molecule of type $m$. The propagation $r_x(t)$ is related to the transmission $x$ through the initial conditions

$$r_x(\tau) = (m, B(\tau)) = (m, 0), \tag{3}$$

since $x = (m, \tau)$ is transmitted from the origin at time $\tau$.

Let $\mathcal{S}$ represent the set of relevant physical states of a molecule (e.g., velocity) that the receiver is capable of measuring. A *reception* $y$ is defined as the triple

$$y = (m, s, \tau') \in \mathcal{M} \times \mathcal{S} \times \mathbb{R}, \tag{4}$$

representing the observation at the receiver of molecule $m \in \mathcal{M}$, with physical state $s \in \mathcal{S}$, at time $\tau' \in \mathbb{R}$. (The need to measure physical states at the receiver is a technical requirement so that the Brownian motion is Markovian under general models of motion.) Recalling our assumption that the receiver only interacts with molecules as they cross the boundary, for each propagation $r_x(t)$, let $\mathcal{T}_{r_x(t)}$ represent the set of time instants such that

$$\mathcal{T}_{r_x(t)} = \{\tau_1', \tau_2', \ldots\} = \{\tau' : B(\tau') = d\}, \tag{5}$$

that is, the set of time instants such that the molecule is located at the boundary. Let $\mathcal{Y}_{r_x(t)}$ represent the set of receptions corresponding to the propagation $r_x(t)$. Then

$$\mathcal{Y}_{r_x(t)} = \{y_0, y_1, \ldots\} = \{y : \tau' \in \mathcal{T}_{r_x(t)}, y = (m, s, \tau')\} \tag{6}$$

is the set of receptions corresponding to $\mathcal{T}_{r_x(t)}$. Further, $\mathcal{Y}_{r_x(t)}$ is the set of observations corresponding to the transmission $x$.

In the sequel, we will represent the sequences of transmissions and receptions of all molecules as vectors, writing $\mathbf{x} = [x_1, x_2, \ldots]$ and $\mathbf{y} = [y_1, y_2, \ldots]$ as vectors of transmissions and receptions, respectively.

### B. Ideal model for transmitter and receiver

In the previous section, we assumed that the molecules within the channel obeyed three assumptions, which we will justify later as physically realistic. However, to model the transmitter and receiver, we take a different approach: we use a mathematically convenient model, and show that this model is *ideal* in the sense that relaxing these assumptions leads to systems with lower mutual information. As a result, we obtain a model that is simple, that abstracts away the details of chemical processing at the terminals, and that is a valid upper bound on performance for any implementation of transmitter and receiver.

We define three properties of a general *ideal model* for molecular communication, as follows:

1) **No control error at the transmitter.** At the transmitter, we assume that the relevant properties of every transmitted molecule can be controlled exactly.

2) **No measurement error at the receiver.** At the receiver, we assume that the relevant properties of every received molecule can be measured exactly.

3) **The receiver absorbs arriving molecules.** At the first passage time of each molecule, the receiver removes the molecule from the system.

It is obvious that the first two properties are ideal, and in the following result we prove that the third property is ideal. Define this property formally as follows: partition the vector of receptions into $[\mathbf{y}^{(1)}, \mathbf{y}^{(2)}]$, where $\mathbf{y}^{(1)}$ contains the receptions corresponding to the first passage times at the receiver, and $\mathbf{y}^{(2)}$ contains all other receptions. That is, for each element $y_i^{(1)} = (m_i(\tau_i), s, \tau_i')$, there is a propagation $r_x(t)$ such that $\tau_i' = \min \mathcal{T}_{r_x(t)}$, and vice versa. Then $\mathbf{y} := \mathbf{y}^{(1)}$, and the receiver discards $\mathbf{y}^{(2)}$. (Equivalently, each propagation $r_x(t)$ is terminated after its first passage time, so $\mathbf{y}^{(2)}$ is always empty). Then we have the following:

*Theorem 1:* Let $I(\mathbf{X}; \mathbf{Y})$ represent mutual information of a molecular communication system under the ideal model. Then $I(\mathbf{X}; \mathbf{Y}) = I(\mathbf{X}; \mathbf{Y}^{(1)}) = I(\mathbf{X}; \mathbf{Y}^{(1)}, \mathbf{Y}^{(2)})$. Further, letting $\mathbf{y}'$ represent $[\mathbf{y}^{(1)}, \mathbf{y}^{(2)}]$, sorted in order of arrival time (i.e., without prior knowledge of which arrivals are first arrivals), then $I(\mathbf{X}; \mathbf{Y}) \geq I(\mathbf{X}; \mathbf{Y}')$.

*Proof:* We can first write

$$
\begin{aligned}
I(\mathbf{X}; \mathbf{Y}^{(1)}, \mathbf{Y}^{(2)}) &= H(\mathbf{Y}^{(2)} \,|\, \mathbf{Y}^{(1)}) + H(\mathbf{Y}^{(1)}) + H(\mathbf{X}) - H(\mathbf{Y}^{(2)} \,|\, \mathbf{Y}^{(1)}, \mathbf{X}) - H(\mathbf{Y}^{(1)}, \mathbf{X}) \quad (7) \\
&= H(\mathbf{Y}^{(2)} \,|\, \mathbf{Y}^{(1)}) + H(\mathbf{Y}^{(1)}) + H(\mathbf{X}) - H(\mathbf{Y}^{(2)} \,|\, \mathbf{Y}^{(1)}) - H(\mathbf{Y}^{(1)}, \mathbf{X}) \quad (8) \\
&= I(\mathbf{X}; \mathbf{Y}^{(1)}), \quad (9)
\end{aligned}
$$

where the second equality follows since the Brownian motion is Markovian, by assumption. However, since $\mathbf{y}^{(1)} = \mathbf{y}$, we have that

$$
I(\mathbf{X}; \mathbf{Y}) = I(\mathbf{X}; \mathbf{Y}^{(1)}) = I(\mathbf{X}; \mathbf{Y}^{(1)}, \mathbf{Y}^{(2)}). \quad (10)
$$

Since $\mathbf{y}'$ is formed by processing $[\mathbf{y}^{(1)}, \mathbf{y}^{(2)}]$, $I(\mathbf{X}; \mathbf{Y}^{(1)}, \mathbf{Y}^{(2)}) \geq I(\mathbf{X}; \mathbf{Y}')$ follows from the data processing inequality, and so the theorem follows. ∎

Intuitively, the theorem states that all the information in the molecules is in the first arrivals $\mathbf{y}^{(1)}$. If the receiver knows which are the first arrivals (e.g., by removing a molecule after its first arrival, or by somehow "tagging" the molecule as having been observed already), it can safely ignore any subsequent arrivals of the same molecule. Meanwhile, taking away the receiver's knowledge of which are the first arrivals can only hurt the mutual information.

We give the following example of a system that violates one of the ideal model assumptions, but that is nonetheless practically important, and will be used in subsequent sections:

*Example 1 (Counting detector):* In the counting detector, time is partitioned into segments of length $T$. The detector forms a sequence $\mathbf{C} = [c_1, c_2, \ldots]$ of integers, where $c_j$ represents the count of arriving molecules on the interval $[jT, (j+1)T)$. (If there is more than one type of molecule in the system, $c_j$ represents the counts of each type of molecule). More formally,

$$
c_j = |\{y_j : y_j = (m_j, s_j, \tau_j'), jT \leq \tau_j' < (j+1)T\}|. \quad (11)
$$

As in the ideal case, we assume that the transmitter is error-free, and that the receiver absorbs arriving molecules. However, the measurements taken by the receiver contain quantization error, since the arrival times are quantized to the intervals $[jT, (j+1)T)$. Thus, it is clear that the counting detector has lower mutual information than the ideal model. *(End of example.)*

## C. Statistical model of the channel

Consider a system employing the ideal model from Section II-B, and suppose a single molecule is transmitted. Then if $x = (m, \tau)$, the molecule will arrive at the receiver as

$$
y = (m, x, \tau') = (m, s, \tau + n_t), \quad (12)
$$

where $n_t$ is the first passage time at the receiver for the Brownian motion. Further, from Theorem 1, measurements taken at the first passage time contain all the relevant information that the receiver needs about the molecule.

Assuming that information is carried in the release time $\tau$, and disregarding the state $s$, the first passage time $n_t$ may be viewed as additive noise. In particular, suppose $\mathbf{x} \in (\mathcal{M} \times \mathbb{R})^n$ is a vector of transmissions, and suppose $\mathbf{y} \in (\mathcal{M} \times \mathcal{S} \times \mathbb{R})^n$ is the corresponding vector of receptions, considering only the first passage times at the receiver. For each element $x_i = (m_i, \tau_i)$ of $\mathbf{X}$, there is a corresponding reception $y_j = (m_i, s_i, \tau_i + n_{t,i})$, with corresponding first passage time $n_{t,i}$, of $\mathbf{Y}$. However, it is not necessarily true that $i = j$, as the random delays $n_{t,i}$ may cause the molecules to arrive out of order, and the receiver does not generally know the order of transmission. Letting $\mathbf{u} = [u_1, u_2, \ldots]$, where $u_i = (m_i, s_i, t_i + n_{t,i})$, we generally observe

$$\mathbf{y} = \mathrm{sort}_\tau(\mathbf{u}), \tag{13}$$

where the function $\mathrm{sort}_\tau : (\mathcal{M} \times \mathcal{S} \times \mathbb{R})^n \to (\mathcal{M} \times \mathcal{S} \times \mathbb{R})^n$ sorts a vector of receptions in increasing order of arrival time. As a result, $\mathbf{y}$ are the *order statistics* of $\mathbf{u}$, sorted with respect to $\tau_i'$.

If $\mathbf{w}$ is an $n$-dimensional vector random variable with independent and non-identically-distributed elements, and $\mathbf{z}$ are the order statistics of $\mathbf{w}$, then the Bapat-Beg theorem [18, Theorem 4.1] states that

$$\Pr(Z_1 < z_1, Z_2 < z_2, \ldots, Z_n < z_n) =$$
$$\mathrm{per} \left( \begin{bmatrix} F_{W_1}(z_1) & F_{W_1}(z_2) - F_{W_1}(z_1) & \cdots & F_{W_1}(z_n) - F_{W_1}(z_{n-1}) \\ F_{W_2}(z_1) & F_{W_2}(z_2) - F_{W_2}(z_1) & \cdots & F_{W_2}(z_n) - F_{W_2}(z_{n-1}) \\ \vdots & \vdots & \ddots & \vdots \\ F_{W_n}(z_1) & F_{W_n}(z_2) - F_{W_n}(z_1) & \cdots & F_{W_n}(z_n) - F_{W_n}(z_{n-1}) \end{bmatrix} \right), \tag{14}$$

where $F_{W_i}(\cdot)$ is the cumulative distribution function (CDF) for the $i$th element of $\mathbf{w}$, and where $\mathrm{per}(\cdot)$ represents the *permanent* of a matrix [19]. For an $n \times n$ matrix $\mathbf{M} = [m_{ij}]$, $\mathrm{per}(\mathbf{M})$ is given by

$$\mathrm{per}(\mathbf{M}) = \sum_{\pi \in \mathcal{P}_n} \prod_{i=1}^n m_{i\pi(i)}, \tag{15}$$

where $\mathcal{P}_n$ is the set of all permutations of $\{1, 2, \ldots, n\}$. For our system, we show in Appendix A that the probability density function (PDF) $f_{\mathbf{Y}|\mathbf{X}}(\mathbf{y}|\mathbf{x})$ is given by

$$f_{\mathbf{Y}|\mathbf{X}}(\mathbf{y}|\mathbf{x}) = \mathrm{per} \left( \begin{bmatrix} f_{Y_1|X_1}(y_1|x_1) & f_{Y_2|X_1}(y_2|x_1) & \cdots & f_{Y_n|X_1}(y_n|x_1) \\ f_{Y_1|X_2}(y_1|x_2) & f_{Y_2|X_2}(y_2|x_2) & \cdots & f_{Y_n|X_2}(y_n|x_2) \\ \vdots & \vdots & \ddots & \vdots \\ f_{Y_1|X_n}(y_1|x_n) & f_{Y_2|X_n}(y_2|x_n) & \cdots & f_{Y_n|X_n}(y_n|x_n) \end{bmatrix} \right), \tag{16}$$

where $f_{Y_i|X_j}(y_i|x_j)$ represents the probability that arrival $y_i$ corresponds to the transmission $x_j$. If $y_i = (m_i, s_i, \tau'_i)$ and $x_j = (m_j, \tau_j)$, then we have that

$$
\begin{aligned}
f_{Y_i|X_j}(y_i|x_j) &= f_{M_j|M_i}(m_j|m_i) f_{S_i, T'_i|T_j}(s_i, \tau'_i|\tau_j) \\
&= \Delta(m_j, m_i) f_{S_i, T'_i|T_j}(s_i, \tau'_i|\tau_j),
\end{aligned}
\tag{17}
$$

where $\Delta(m_j, m_i)$ represents the Kronecker delta function, which appears since (by assumption) molecules do not change in transit. From (16) and (17), $f_{S_i, T'_i|T_j}(s_i, \tau'_i|\tau_j)$ is sufficient to completely characterize the system.

In spite of its superficial similarity between the form of (15) and the calculation of the determinant, the permanent is a member of a class of problems, known as #P-complete,[1] which are believed to be intractable [20]. The fastest known algorithm for calculating the permanent of an $n \times n$ matrix, given in [21], has $\Theta(n2^n)$ complexity; while most bounds and approximations have either low accuracy or relatively high complexity (e.g., [22]). Obviously, these facts significantly complicate the calculation of the information rate of this channel. Although exact calculation of the information rate appears to be difficult for any large system, in later sections we will give approaches for bounding the information rate.

### D. Separation into parallel channels for distinguishable molecules

We now give a useful result that simplifies our analysis in the remainder of the paper: namely, that distinguishable molecules can be separated into independent parallel channels.

For each $i \in \mathcal{M}$, let $\mathbf{x}^{(i)}$ and $\mathbf{y}^{(i)}$ represent the vectors of transmissions and receptions, respectively, corresponding to molecule $i$. Then we have the following:

*Theorem 2:* $f_{\mathbf{Y}|\mathbf{X}}(\mathbf{y}|\mathbf{x}) = \prod_{i \in \mathcal{M}} f_{\mathbf{Y}^{(i)}|\mathbf{X}^{(i)}}(\mathbf{y}^{(i)} \mid \mathbf{x}^{(i)})$ .

*Proof:* Let $\mathcal{M} = \{m_1, m_2, \ldots, m_{|\mathcal{M}|}\}$. Suppose $\mathbf{x}$ and $\mathbf{y}$ are rearranged so that

$$
\mathbf{x} = \left[ \mathbf{x}^{(m_1)}, \mathbf{x}^{(m_2)}, \ldots, \mathbf{x}^{(m_{|\mathcal{M}|})} \right],
\tag{18}
$$

and

$$
\mathbf{y} = \left[ \mathbf{y}^{(m_1)}, \mathbf{y}^{(m_2)}, \ldots, \mathbf{y}^{(m_{|\mathcal{M}|})} \right].
\tag{19}
$$

Since row and column permutations do not affect the permanent, this rearrangement does not affect $f_{\mathbf{Y}|\mathbf{X}}(\mathbf{y}|\mathbf{x})$. Let $\mathbf{H} = [h_{i,j}]$, where $h_{i,j} = f_{Y_i|X_j}(y_i|x_j)$; then from (16), $f_{\mathbf{Y}|\mathbf{X}}(\mathbf{y}|\mathbf{x}) = \mathrm{per}(\mathbf{H})$. Furthermore, for each $m_a, m_b \in \mathcal{M}$, let $\mathbf{H}^{(m_a, m_b)} = [h_{i,j}^{(m_a, m_b)}]$, where $h_{i,j}^{(m_a, m_b)} = f_{Y_i|X_j}(y_i^{(m_a)}|x_j^{(m_b)})$. Then, given the

---

[1] #P-complete is pronounced "sharp-P complete".

rearrangement in (18)-(19),

$$\mathbf{H} = \begin{bmatrix} \mathbf{H}^{(m_1,m_1)} & \mathbf{H}^{(m_1,m_2)} & \dots & \mathbf{H}^{(m_1,m_{|\mathcal{M}|})} \\ \mathbf{H}^{(m_2,m_1)} & \mathbf{H}^{(m_2,m_2)} & \dots & \mathbf{H}^{(m_2,m_{|\mathcal{M}|})} \\ \vdots & \vdots & \ddots & \vdots \\ \mathbf{H}^{(m_{|\mathcal{M}|},m_1)} & \mathbf{H}^{(m_{|\mathcal{M}|},m_2)} & \dots & \mathbf{H}^{(m_{|\mathcal{M}|},m_{|\mathcal{M}|})} \end{bmatrix} \tag{20}$$

$$= \begin{bmatrix} \mathbf{H}^{(m_1,m_1)} & \mathbf{0} & \cdots & \mathbf{0} \\ \mathbf{0} & \mathbf{H}^{(m_2,m_2)} & \cdots & \mathbf{0} \\ \vdots & \vdots & \ddots & \vdots \\ \mathbf{0} & \mathbf{0} & \cdots & \mathbf{H}^{(m_{|\mathcal{M}|},m_{|\mathcal{M}|})} \end{bmatrix}, \tag{21}$$

where $\mathbf{0}$ represents an all-zero matrix of the appropriate dimension; the all-zero matrices in the second equality follow from the delta function in (17). Since $\mathbf{H}$ is a block-diagonal matrix, a property of the permanent is that [23]

$$\mathrm{per}(\mathbf{H}) = \prod_{i \in \mathcal{M}} \mathrm{per}\left(\mathbf{H}^{(i,i)}\right), \tag{22}$$

and the theorem straightforwardly follows. ∎

*Example 2:* Suppose that every molecule is distinguishable, that is, there is exactly one element each in $\mathbf{x}^{(m_1)}, \mathbf{x}^{(m_2)}, \dots, \mathbf{x}^{(m_{|\mathcal{M}|})}$ and $\mathbf{y}^{(m_1)}, \mathbf{y}^{(m_2)}, \dots, \mathbf{y}^{(m_{|\mathcal{M}|})}$. Then (21) becomes

$$\mathbf{H} = \begin{bmatrix} f_{Y|X}(y^{(m_1)}|x^{(m_1)}) & 0 & \cdots & 0 \\ 0 & f_{Y|X}(y^{(m_2)}|x^{(m_2)}) & \cdots & 0 \\ \vdots & \vdots & \ddots & \vdots \\ 0 & 0 & \cdots & f_{Y|X}(y^{(m_{|\mathcal{M}|})}|x^{(m_{|\mathcal{M}|})}) \end{bmatrix}, \tag{23}$$

dropping the boldface notation from $x^{(m_i)}$ and $y^{(m_i)}$, as they are scalars in this example. Since (23) is a diagonal matrix, it can be easily verified that

$$f_{\mathbf{Y}|\mathbf{X}}(\mathbf{y} \mid \mathbf{x}) = \mathrm{per}(\mathbf{H}) = \prod_{i=1}^{|\mathcal{M}|} f_{Y|X}(y^{(m_{|\mathcal{M}|})} \mid x^{(m_{|\mathcal{M}|})}), \tag{24}$$

which is equivalent to the input-output PDF of any independent additive noise channel.

*(End of example.)*

As a result of Theorem 2, for any system where $|\mathcal{M}| > 1$, we can separate the problem into independent and identical parallel channels corresponding to each type of molecule in $\mathcal{M}$. Thus, the general information rate problem reduces to the problem of finding information rates when $|\mathcal{M}| = 1$, i.e., when all the molecules in the system are indistinguishable.

*E. The Wiener-Ideal Model*

If the Brownian motion is modeled by the Wiener process [17], then there are no relevant physical states (i.e., $\mathcal{S} = \emptyset$), and it can be shown that

$$f_{N_t}(n_t) = \frac{d}{\sqrt{2\pi\sigma^2 n_t^3}} \exp\left(-\frac{d^2}{2\sigma^2 n_t}\right). \tag{25}$$

Notice from (25) that $f_{N_t}(n_t)$ has a very long tail, which decays with $n_t^{-3/2}$; neither the mean nor the variance of this distribution are finite. The two parameters of the distribution are $\sigma^2$, the intensity of the Brownian motion; and $d$, the distance from the transmitter to the receiver. There exist other models for Brownian motion, such as the Ornstein-Uhlenbeck process, where the only relevant physical state is velocity (i.e., $\mathcal{S} = \mathbb{R}$); but in that case (and in most cases other than the Wiener process), $f_{N_t}(n_t)$ cannot be expressed in closed form.

Suppose the ideal model from Section II-B is used. If the Brownian motion is modeled by the Wiener process, we can simplify (17) to

$$f_{Y_i|X_j}(y_i|x_j) = \Delta(m_i, m_j) f_{N_t}(\tau_i' - \tau_j). \tag{26}$$

Further, bearing in mind Theorem 2, for each $m \in \mathcal{M}$, we can disregard the $\Delta(m_i, m_j)$ term and assume that $m$ is the only type of molecule in the channel. We can now write

$$f_{Y_i|X_j}(y_i|x_j) = f_{N_t}(\tau_i' - \tau_j), \tag{27}$$

so that the system is fully characterized by the first passage time distribution. Thus, the transmissions are all of the form $x = (m, \tau)$, and the receptions are all of the form $y = (m, , \tau')$ (noting the blank in place of $s$, since $\mathcal{S} = \emptyset$); the only task of the transmitter is to set the departure time $\tau$, and the only task of the receiver is to measure the first passage time $\tau'$. We call this straightforward model the *Wiener-Ideal* (WI) model.

We now consider whether our modeling framework in general, and the WI model in particular, are physically realistic. We began with three fundamental assumptions – that Brownian motion is Markovian, that the motions are statistically independent, and that molecules don't change in transit. The first two assumptions are common in the diffusion literature (e.g., see [17]). The third assumption depends on the type of molecule in use, but there certainly exist molecules that are stable over the relatively short time scales we consider. Furthermore, it is appropriate to assume that the motions are statistically described by the Wiener process, so long as the Brownian motion is nearly free of friction [24]. If friction is significant, this assumption can be relaxed by substituting the first arrival time distribution for the Wiener process with the appropriate distribution; none of our subsequent methods depend specifically on the Wiener process. Finally, we have the ideal modeling assumptions for the transmitter and receiver, which are less realistic, but which we proposed as a way of deliberately trading off realism in order to obtain abstraction. Further, as we saw in Theorem 1, these modeling assumptions are meaningful in an information-theoretic sense. As a result, we may conclude that the WI model is physically realistic, and for the remainder of the paper, we will deal exclusively with this model.

## III. Some simplified views of information rate

Under the WI model, we showed that $f_{\mathbf{Y}|\mathbf{X}}(\mathbf{y}|\mathbf{x})$ is intractable for a realistic number of molecules. However, we can nonetheless give some simplified calculations of capacity, which will demonstrate the importance of constraining the input distribution. Note that in this section, and for the remainder of the paper, logarithms are assumed to be base 2 unless otherwise stated.

We assume the WI model, but the results are valid for any first passage time distribution where $f_{N_t}(n_t) > 0$ for any $n_t > 0$, and where $\lim_{n_t \to \infty} \int_0^{n_t} f_N(\nu) d\nu = 1$ (i.e., the molecule arrives in finite time with probability 1). Let $\mathbf{x}$ and $\mathbf{y}$ represent $n$-fold vectors of transmissions and receptions, let $f_X(\mathbf{x})$ represent the input distribution, and let $T_S$ represent the total observation time for the system. Then we can state the following result.

*Theorem 3:* For any $n$,

$$\lim_{T_S \to \infty} \max_{f_{\mathbf{X}}(\mathbf{x})} \frac{I(\mathbf{X}; \mathbf{Y})}{n} = \infty. \tag{28}$$

Further, for any $T_S$,

$$\lim_{n \to \infty} \max_{f_{\mathbf{X}}(\mathbf{x})} \frac{I(\mathbf{X}; \mathbf{Y})}{T_S} = \infty. \tag{29}$$

*Proof:* In both cases, the statement is proved by finding special cases in which the information rate is infinite.

To prove the first statement, suppose $T_S$ is partitioned into intervals of duration $\log T_S$, and suppose $f_{\mathbf{X}}(\mathbf{x})$ is chosen so that all $n$ molecules are released at the beginning of a single interval. As $T_S \to \infty$, $\log T_S \to \infty$, so the probability that the molecules arrive in the same interval is 1 (by assumption). Using this method, $\log \log T_S$ bits can be transmitted without error. Since $\log \log T_S \to \infty$ as $T_S \to \infty$, then for each $n$, $\lim_{T_S \to \infty} \max_{f_{\mathbf{X}}(\mathbf{x})} I(\mathbf{X}; \mathbf{Y})/n = \infty$.

To prove the second statement, we take the same strategy as in [15] for infinite-server queues, which we restate here for completeness. Divide $T_S$ into intervals of size $\nu$, where $f_{\mathbf{X}}(\mathbf{x})$ is again chosen so as to release all $n$ molecules at the beginning of this interval. The receiver's strategy is to wait for the first molecule to arrive, and decide that it was transmitted in the same interval in which it arrived; all remaining molecules are ignored. Since $f_N(n) > 0$ for all $n > 0$ (under the WI model), as $n \to \infty$ the probability of error in this strategy goes to zero. Thus, $\log T_S/\nu$ bits can be transmitted using this strategy, and as $\nu \to 0$, $\log T_S/\nu \to \infty$. Thus, for any $T_S$, $\lim_{n \to \infty} \max_{f_{\mathbf{X}}(\mathbf{x})} I(\mathbf{X}; \mathbf{Y})/T_S = \infty$ . ∎

The strategies to achieve these two results – waiting infinitely long and releasing an infinite number of molecules at once, respectively – are impractical because both time and molecules are costly resources in molecular communication. Furthermore, physical limitations (such as saturation) may prevent these strategies. However, Theorem 3 does give us some intuition concerning the behavior of our system in practical settings: for example, we expect that the number of bits per molecule should be high when molecules are sparse. In the remainder of the paper, we attempt to find information rates per unit time, or per molecule, given appropriate constraints on the input distribution.

## IV. Achievable lower bounds on information rate

### A. Bounds from approximate distributions

We know that mutual information can be written

$$I(X;Y) = \lim_{T \to \infty} \frac{1}{T} E \left[ \log \frac{f_{\mathbf{Y}|\mathbf{X}}(\mathbf{y} \mid \mathbf{x})}{f_{\mathbf{Y}}(\mathbf{y})} \right], \tag{30}$$

so long as the limit exists. Furthermore, it is easy to generate instances of $\mathbf{x}$ and $\mathbf{y}$, so any expectation of these variables may be tractably obtained using *Monte Carlo* methods. Unfortunately, in (30), the function $f_{\mathbf{Y}|\mathbf{X}}(\mathbf{y} \mid \mathbf{x})/f_{\mathbf{Y}}(\mathbf{y})$ is itself intractable.

Instead, suppose that there exist tractable approximations $g(\mathbf{y} \mid \mathbf{x})$, and $g(\mathbf{y})$ for $f_{\mathbf{Y}|\mathbf{X}}(\mathbf{y} \mid \mathbf{x})$ and $f_{\mathbf{Y}}(\mathbf{y})$, respectively, which have the following properties:

1) $\int_{\mathbf{y}} g(\mathbf{y} \mid \mathbf{x}) = 1$ and $g(\mathbf{y} \mid \mathbf{x}) \geq 0$ for all $\mathbf{x}, \mathbf{y}$ (i.e., $g(\mathbf{y} \mid \mathbf{x})$ is a valid probability density function); and

2) Given the true input distribution $f_{\mathbf{X}}(\mathbf{x})$, $g(\mathbf{y})$ is found by $\int_{\mathbf{x}} g(\mathbf{y} \mid \mathbf{x}) f_{\mathbf{X}}(\mathbf{x})$.

If the approximations $g(\mathbf{y} \mid \mathbf{x})$ and $g(\mathbf{y})$ are sufficiently good, a close approximation to $I(X;Y)$ might be found by substituting $g(\mathbf{y} \mid \mathbf{x})/g(\mathbf{y})$ in place of $f_{\mathbf{Y}|\mathbf{X}}(\mathbf{y} \mid \mathbf{x})/f_{\mathbf{Y}}(\mathbf{y})$ in (30) and using *Monte Carlo* expectation.

In fact, the approximation is a lower bound on the true $I(X;Y)$. The following result was proved in [16] and elsewhere, which we restate and prove here:

*Theorem 4:* Given $g(\mathbf{y} \mid \mathbf{x})$ and $g(\mathbf{y})$ defined as above,

$$I(X;Y) \geq \lim_{T \to \infty} \frac{1}{T} E \left[ \log \frac{g(\mathbf{y} \mid \mathbf{x})}{g(\mathbf{y})} \right], \tag{31}$$

where the expectation is taken with respect to the true distribution of $\mathbf{x}$ and $\mathbf{y}$.

*Proof:* Let $g(\mathbf{x} \mid \mathbf{y}) = g(\mathbf{y} \mid \mathbf{x}) f_{\mathbf{X}}(\mathbf{x})/g(\mathbf{y})$. We can rewrite the expectation in (31) as

$$
\begin{aligned}
E\left[\log \frac{g(\mathbf{y} \mid \mathbf{x})}{g(\mathbf{y})}\right] &= E\left[\log \frac{g(\mathbf{y} \mid \mathbf{x}) f_{\mathbf{X}}(\mathbf{x})}{g(\mathbf{y}) f_{\mathbf{X}}(\mathbf{x})}\right] \\
&= H(\mathbf{X}) + E\left[\log g(\mathbf{x} \mid \mathbf{y})\right] \\
&= H(\mathbf{X}) - H(\mathbf{X} \mid \mathbf{Y}) - D\left(f_{\mathbf{X}|\mathbf{Y}}(\mathbf{x}|\mathbf{y}) \parallel g(\mathbf{x}|\mathbf{y})\right) \\
&= I(\mathbf{X}; \mathbf{Y}) - D\left(f_{\mathbf{X}|\mathbf{Y}}(\mathbf{x}|\mathbf{y}) \parallel g(\mathbf{x}|\mathbf{y})\right),
\end{aligned}
\tag{32}
$$

where $D(f \parallel g)$ represents Kullback-Leibler (KL) divergence. The theorem immediately follows from (32) and the properties of KL divergence. ∎

The bound in (31) has the interesting physical interpretation as an achievable rate for a decoder that assumes that $g(\mathbf{y} \mid \mathbf{x})$ is the correct input-output distribution. As a result, the bound in (31) is an achievable bound, in that we could (in principle) construct a device to achieve reliable communication at the rate given by the bound.

### B. The trapdoor channel and trivial approximations

As we see from Theorem 4, the tightness of the bound is governed by a term related to the Kullback-Leibler divergence between the approximation and the true distribution, so better approximations will lead to better bounds. Our challenge is to find good tractable approximations $g(\mathbf{y}|\mathbf{x})$, but as we point out in this section, seemingly good candidates for $g(\mathbf{y}|\mathbf{x})$ can lead to trivial bounds.

We can write

$$
E\left[\log \frac{g(\mathbf{y} \mid \mathbf{x})}{g(\mathbf{y})}\right] = \int_{\mathbf{x},\mathbf{y}} f_{\mathbf{X},\mathbf{Y}}(\mathbf{x},\mathbf{y}) \log \frac{g(\mathbf{y} \mid \mathbf{x})}{g(\mathbf{y})}.
\tag{33}
$$

If there exist $\mathbf{x}$ and $\mathbf{y}$ such that $g(\mathbf{y} \mid \mathbf{x}) = 0$, while $f_{\mathbf{X},\mathbf{Y}}(\mathbf{x},\mathbf{y}) > 0$ and $g(\mathbf{y}) > 0$, we refer to $g(\mathbf{y} \mid \mathbf{x})$ as a *trivial approximation*. For any trivial approximation, it is easy to see that the expectation in (33) returns a value of $-\infty$. A sufficient condition to avoid a trivial approximation is to require $g(\mathbf{y} \mid \mathbf{x}) > 0$ for all $\mathbf{x}$ and $\mathbf{y}$, and we will require this condition to hold for all candidate approximations in the remainder of the paper.

The reader may believe that the trapdoor (or "chemical") channel [12] is a promising candidate for $g(\mathbf{y} \mid \mathbf{x})$; this channel is described as follows. Let $\mathcal{U}$ represent a set of symbols, and using time index $t$, let $u_t \in \mathcal{U}$ and $v_t \in \mathcal{U}$ represent inputs and outputs, respectively; and let the multiset $\mathcal{S}_t = \{s_1, s_2, \ldots, s_{|\mathcal{S}|}\} \in \mathcal{U}^n$ represent the channel state. At time $t$, $u_t$ is provided to the channel, and $v_t$ is selected uniformly at random from the multiset $\{u_t\} \cup \mathcal{S}_t$. (The selection probability can be generalized to something non-uniform, but this does not change our subsequent analysis.) Finally, we set $\mathcal{S}_{t+1} = (\{u_t\} \cup \mathcal{S}_t) \backslash v_t$. This is commonly likened to a bag of billiard balls, where $\mathcal{S}_t$ represents the balls already in the bag, $x_t$ represents a ball dropped into the bag, and $v_t$ represents a ball removed from the bag. This process also has a superficial similarity to chemical diffusion, which has been noted by some authors [13].

Unfortunately, this channel model does not satisfy the sufficient condition we gave to avoid trivial approximations. To see this, let $\mathbf{u} = [u_1, u_2, \ldots, u_{|\mathcal{S}|+1}]$ and $\mathbf{v} = [v_1, v_2, \ldots, v_{|\mathcal{S}|+1}]$ be length-$(|\mathcal{S}| + 1)$ sequences of trapdoor channel inputs and outputs, respectively. Say $u_1 = u_2 = \ldots = u_{|\mathcal{S}|+1}$, and $v_1 = v_2 = \ldots = v_{|\mathcal{S}|+1}$, but $u_i \neq v_i$ for any $i = 1, 2, \ldots, |\mathcal{S}| + 1$; in the billiard ball analogy, $|\mathcal{S}| + 1$ blue balls are dropped into the bag, and $|\mathcal{S}| + 1$ red balls are removed. Since the largest number of red balls initially in the bag is $|\mathcal{S}|$, and no additional red balls are inserted, this scenario is clearly impossible – so $f_{\mathbf{V}|\mathbf{U}}(\mathbf{v}|\mathbf{u}) = 0$.

There exist a few obvious mappings from $\mathbf{u}$ to $\mathbf{x}$ and from $\mathbf{v}$ to $\mathbf{y}$: for example, a red ball may represent a transmission or reception at the input or output, respectively; and a blue ball may represent no transmission or reception. However, for this mapping, there is no way to structure the input distribution so as to avoid a trivial approximation: an imbalance between transmissions and receptions, which exceeds $|\mathcal{S}|$, is impossible in the trapdoor channel. Further, using the WI model, it is straightforward to show for most input distributions that the expected number of molecules in transit is infinite, so it is generally possible to find sequences $\mathbf{x}$ and $\mathbf{y}$ with $f_{\mathbf{X},\mathbf{Y}}(\mathbf{x}, \mathbf{y}) > 0$ for the true distribution, but $g(\mathbf{y}|\mathbf{x}) = 0$ under the finite-state trapdoor model. As a result, it is not easy to see how to avoid trivial approximations with the trapdoor channel, and so we do not consider that model any further in this paper.

### C. A Simple Approximate Model

Bearing the previous section in mind, we need to find a model which provides a reasonably good, yet tractable, pair of approximations $g(\mathbf{y}|\mathbf{x})$ and $g(\mathbf{y})$. Our goal in this section is to introduce a simple model that leads to useful lower bounds, which we generalize in the next section.

The basis of our simple approximate model is the counting detector of Example 1. As in that example, let $\mathbf{c} = [c_1, c_2, \ldots]$ represent the counts in each interval. For each transmission in $\mathbf{x} = [x_1, x_2, \ldots]$, let $a_j = 1$ if the transmission $x_j$ arrives during the interval $[jT, (j+1)T)$, and $a_j = 0$ otherwise. Then the number of arrivals can be written

$$c_j = \sum_{j=1}^{n} a_j, \tag{34}$$

Accordingly, we are actually approximating $f_{\mathbf{C}|\mathbf{X}}(\mathbf{c}|\mathbf{x})$, the conditional probability of count vector $\mathbf{c}$ given transmissions $\mathbf{x}$, with $g(\mathbf{c}|\mathbf{x})$. By Theorem 4, this approximation provides a lower bound on the mutual information of the counting detector, while by Example 1, the mutual information of the counting detector provides a lower bound on the mutual information of the WI model.

TABLE I

ILLUSTRATION OF THE ACCURACY OF THE POISSON APPROXIMATION FOR BACKGROUND ARRIVALS; $10^4$ TRIALS.

| k | 0 | 1 | 2 | 3 | 4 | >= 5 |
|---------|--------|--------|--------|--------|--------|-------------------------|
| Actual | 0.6921 | 0.2602 | 0.0423 | 0.0047 | 0.0007 | 0 |
| Poisson | 0.6965 | 0.2519 | 0.0456 | 0.0055 | 0.0005 | $3.8215 \cdot 10^{-5}$ |

For convenience, we constrain the input distribution so that transmissions only occur on the boundary between intervals, i.e., only at times $jT$, for integer $j$. Partition $\mathbf{x}$ into subvectors such that

$$\mathbf{x} = [\mathbf{x}_1, \mathbf{x}_2, \ldots], \tag{35}$$

where $\mathbf{x}_j$ is the vector of transmissions that occur at the instant $\tau = jT$. We allow the vector $\mathbf{x}_j$ to be empty, which corresponds to the event that there are no transmissions at time $\tau = jT$.

To achieve computational simplicity in the approximate model, we require that

$$g(\mathbf{c} \mid \mathbf{x}) = \prod_{j=1}^{n} g(c_j \mid \mathbf{x}_j), \tag{36}$$

so $c_i$ and $\mathbf{x}_j$ are *assumed independent* if $i \neq j$. This can be interpreted as follows: for molecules released at time $\tau = jT$, at the beginning of the interval $[jT, (j+1)T)$, the receiver only attempts to detect the transmissions $\mathbf{x}_j$ during that interval; if those $\mathbf{x}_j$ do not arrive during this interval, then those molecules is assumed to be "lost". Thus, in addition to the transmissions, there will also be spurious arrivals composed of these "lost" molecules. These arrivals form part of the sum in (34), and it is known that the sum of independent Bernoulli random variables, whether identically distributed or not, can be approximated with the Poisson distribution [25]. Thus, the Poisson distribution can be used to model the additive "noise" from the lost molecules. In Table I, we give the empirical distribution of the background arrivals, produced after the transmission of $10^4$ molecules under the WI model, alongside a Poisson distribution with the same mean. From the table, we see that the Poisson distribution provides a very good approximation for this process.

We now derive $g(c_j \mid \mathbf{x}_j)$. A molecule released at time $\tau$ arrives between time $\tau$ and $\tau + T$ with probability $p_a$, where

$$p_a = \int_{n_t = 0}^{T} f_{N_t}(n_t). \tag{37}$$

Let the function $\eta(\mathbf{x}_j)$ represent the number of transmissions in $\mathbf{x}_j$. Furthermore, for any integer $k$, let

$$\phi(k; \lambda) = \begin{cases} \frac{\lambda^k e^{-\lambda}}{k!}, & k \geq 0; \\ 0, & k < 0. \end{cases} \tag{38}$$

represent the Poisson distribution with parameter $\lambda$. By assumption in $g(c_j \mid \mathbf{x}_j)$, $\eta(\mathbf{x}_j)$ molecules are transmitted at time $jT$, and $c_j$ arrive on the interval $[jT, (j+1)T)$. Suppose that $k$ of the $\eta(\mathbf{x}_j)$ transmissions arrive, which are distributed Bernoulli with probability $p_a$; then $(c_j - k)$ "lost" molecules arrive, which are distributed Poisson. Thus, we have

$$g(c_j \mid \mathbf{x}_j) = \sum_{k=0}^{\eta(\mathbf{x}_j)} \binom{\eta(\mathbf{x}_j)}{k} p_a^k (1-p_a)^{\eta(\mathbf{x}_j)-k} \phi(c_j - k; \lambda). \tag{39}$$

One may then find $g(\mathbf{c})$ by marginalizing (36) over $\mathbf{x}$, in accordance with the necessary properties of $g(\mathbf{c})$. It is obvious that $g(\mathbf{c}|\mathbf{x})$ is a valid probability distribution. Further, from (38) and (39), $g(c_t \mid x_t)$ is nonzero as long as $p_a < 1$, which is true for any $T : 0 < T < \infty$ and any $\lambda : 0 < \lambda < \infty$. Thus, for the given values of $T$ and $\lambda$, this approximation $g(\mathbf{c}|\mathbf{x})$ satisfies the sufficient condition from Section IV-B.

### D. Generalized Approximate Model

We can generalize the technique in the previous section by observing the channel for the arrival of a particular transmission over $iT$ seconds, for some integer $i \geq 1$. As a molecule is transmitted, earlier molecules may still be propagating towards the receiver, and the late arrivals of these propagating molecules are somewhat analogous to inter-symbol interference. Detection is accomplished by assuming a Markov relationship between transmissions and receptions.

Generalizing (37), for nonnegative integers $k$, let $p_{a,k}(\tau)$ represent the probability that a molecule arrives on the interval $[kT, (k+1)T)$, given that it was released at time $\tau$. This probability is given by

$$p_{a,k}(\tau) = \int_{kT-\tau}^{(k+1)T-\tau} f_{N_t}(n_t). \tag{40}$$

We will also use the conditional probability $\bar{p}_{a,j}(\tau)$ of a molecule arriving in the interval $[kT, (k+1)T)$, given that it was released at time $\tau$, and that it did not arrive on the interval $[\tau, kT)$, given by

$$\bar{p}_{a,k}(\tau) = \frac{p_{a,k}(\tau)}{\int_0^{kT} f_{N_t}(n_t)} = \frac{p_{a,k}(\tau)}{\sum_{i=0}^{k-1} p_{a,i}(\tau)}. \tag{41}$$

The state of the channel at any time $t = jT$, written $\mathbf{s}_j$, consists of the transmissions $x$ that were transmitted prior to $jT$, and remain in transit during the interval $[jT, (j+1)T)$. Extending the function $\eta(\mathbf{s}_j)$ to represent the number of transmissions contained in the state $\mathbf{s}_j$, we can write

$$f_{C_j|\mathbf{s}_j,\mathbf{s}_{j+1},\mathbf{x}_j}(c_j \mid \mathbf{s}_j, \mathbf{s}_{j+1}, \mathbf{x}_j) = \begin{cases} 1, & c_j = \eta(\mathbf{s}_j) + \eta(\mathbf{x}_j) - \eta(\mathbf{s}_{j+1}), \\ 0 & \text{otherwise}, \end{cases} \tag{42}$$

since the number of arrived molecules in the $j$th interval is equal to the number already in transit (i.e., $\eta(\mathbf{s}_j)$) and the number added (i.e., $\eta(\mathbf{x}_j)$), minus the number still in transit in the $(j+1)$th interval (i.e., $\eta(\mathbf{s}_{j+1})$) – a sort of "Kirchoff's law" of molecules. Furthermore, since the state sequence $\mathbf{s}_j, \mathbf{s}_{j+1}, \ldots$ only encodes the molecules in transit along with their original transmission time, this sequence is clearly a Markov chain. Refining the $\eta(\cdot)$ function, let $\eta_\tau(\mathbf{x}_t)$ represent the number of transmissions in the vector for which the release time is equal to $\tau$. Further, let $\mathcal{R}$ represent the set of allowed release times. We can then write

$$f_{\mathbf{S}_{j+1}|\mathbf{S}_j,\mathbf{X}_j}(\mathbf{s}_{t+1} \mid \mathbf{s}_t, \mathbf{x}_t) = \prod_{\tau \in \mathcal{R}} \binom{\eta_\tau(\mathbf{s}_j) + \eta_\tau(\mathbf{x}_j)}{\eta_\tau(\mathbf{s}_{j+1})} \bar{p}_{a,j}(\tau)^{\eta_\tau(\mathbf{s}_j)+\eta_\tau(\mathbf{x}_j)-\eta_\tau(\mathbf{s}_{j+1})}(1 - \bar{p}_{a,j}(\tau))^{\eta_\tau(\mathbf{s}_{j+1})}, \quad (43)$$

defining $\binom{a}{b} = 0$ if $b > a$, and $\binom{a}{b} = 1$ if $a = b = 0$. Equation (43) is obtained since each molecule has a certain probability of arrival in the $j$th interval that is dependent on its departure time, so arrivals of molecules transmitted at the same time have the binomial distribution. As a result, letting $\mathbf{S} = [\mathbf{s}_0, \mathbf{s}_1, \ldots]$, we can write

$$f_{\mathbf{C},\mathbf{S}|\mathbf{X}}(\mathbf{c}, \mathbf{S} \mid \mathbf{x}) = f_{\mathbf{S}_0}(\mathbf{s}_0) \prod_{j=0}^{n} f_{C_j|\mathbf{S}_j,\mathbf{S}_{j+1},\mathbf{X}_j}(c_j \mid \mathbf{s}_j, \mathbf{s}_{j+1}, \mathbf{x}_j) f_{\mathbf{S}_{j+1}|\mathbf{S}_j,\mathbf{X}_j}(\mathbf{s}_{j+1} \mid \mathbf{s}_j, \mathbf{x}_j), \quad (44)$$

where $f_{\mathbf{S}_0}(\mathbf{s}_0)$ represents the distribution of the initial channel state.

The statistical model in (44) is not an approximation – it is the true probability of the output and channel state of the counting detector, conditioned on channel input. However, since the state space is generally enormous, there is no tractable way to recover $f_{\mathbf{C}|\mathbf{X}}(\mathbf{c} \mid \mathbf{x})$ from $f_{\mathbf{C},\mathbf{S}|\mathbf{X}}(\mathbf{c}, \mathbf{S} \mid \mathbf{x})$. This should not be surprising, given our derivation of the exact model in Section II-C. However, bearing in mind our approximate model from the previous section, we may constrain the complexity of the state space by *deleting molecules from the state space after a given amount of time.* Similarly to the previous section, these deleted molecules are considered "lost", and their eventual arrivals are assumed to form a Poisson noise process at the receiver. Furthermore, by adjusting the amount of time that molecules are allowed to remain in the state space, it is possible to trade the fidelity of the model against its computational complexity (which scales with the size of the state space).

Once again, for convenience, we assume that transmissions only occur at times $\tau = kT$ for integer values of $k$. However, we constrain the state space so that the state $\mathbf{s}_j$ may only contain transmissions with release times more recent than $(j - i)T$ for some integer $i$. Then we can use

$$\mathcal{R} = \{jT, (j-1)T, (j-2)T, \ldots, (j-i+1)T\} \quad (45)$$

in (43). Meanwhile, $f_{C_j|\mathbf{S}_j,\mathbf{S}_{j+1},\mathbf{X}_j}(c_j \mid \mathbf{s}_j, \mathbf{s}_{j+1}, \mathbf{x}_j)$ must be adjusted to account for the arrivals of molecules that are about to be deleted from the channel state (since these are no longer accounted for in the transition from $\mathbf{s}_t$ to $\mathbf{s}_{t+1}$, as they are in (42)), as well as the Poisson noise process. These issues are dealt with separately. First, we form an intermediate variable $w_j$, which counts the number of arrivals in the absence of the Poisson noise. For this special case, we can simplify the notation: let $\eta_{k,j} = \eta_{(j-k)T}(\mathbf{s}_j)$, and let $\bar{p}_{a,k} = \bar{p}_{a,j}((j-k)T)$, i.e., the probability of arrival in the current interval i.e., the number of molecules

still in transit, and the probability of arrival, respectively, for molecules that were transmitted $k$ intervals ago. Then $w_j$ has PDF

$$f_{W_j|\mathbf{S}_j,\mathbf{S}_{j+1},\mathbf{X}_j}(w_j \mid \mathbf{s}_j, \mathbf{s}_{j+1}, \mathbf{x}_j) =$$
$$\begin{cases} \binom{\eta_{i-1,j}}{r}(\bar{p}_{a,i-1})^{\eta_{i-1,j}-r}(1-\bar{p}_{a,i-1})^r, & w_j = \eta(\mathbf{s}_j) + \eta(\mathbf{x}_j) - \eta(\mathbf{s}_{j+1}) - r \\ & \quad \text{for } 0 \leq r \leq \eta_{i-1,j}, \\ 0, & \text{otherwise.} \end{cases} \tag{46}$$

Finally, given $w_j$, $c_j - w_j$ is distributed Poisson with intensity $\lambda$, with probability given in (38), and the probability of $c_j$ is given by marginalizing with respect to $w_j$, as follows:

$$f_{C_j|\mathbf{S}_j,\mathbf{S}_{j+1},\mathbf{X}_j}(c_j \mid \mathbf{s}_j, \mathbf{s}_{j+1}, \mathbf{x}_j) = \sum_{w_j} f_{W_j|\mathbf{S}_j,\mathbf{S}_{j+1},\mathbf{X}_j}(w_j \mid \mathbf{s}_j, \mathbf{s}_{j+1}, \mathbf{x}_j)\phi(c_j - w_j; \lambda). \tag{47}$$

It can be shown that this is not a trivial approximation, using a similar argument as for the simplified method in the previous section.

This method gives a *sequence of lower bounds* on mutual information, where the bounds are enumerated with respect to $i$; furthermore, $i = 1$ corresponds to the simple "memoryless" approximation from Section IV-C. With appropriate selection of $\lambda$ (which we discuss in Section VI), we conjecture that the value of the lower bound is increasing in $i$, since the fidelity of the approximate model increases as $i$ increases. Furthermore, so long as $\lambda \to 0$ as $i \to \infty$, the approximate output distribution (46) approaches the true distribution (42), while no molecules are "lost" from the channel state, so this approximation is asymptotically correct for the counting detector. Information rates found using this approximate model are given in Section VI.

## V. Upper Bounds on Information Rate

As with the lower bounds, a sequence of straightforward upper bounds are also available, which are generated by providing side information to the decoder that makes the problem tractable.

Consider a channel model that is similar to the WI model, but which operates in the following way: the channel first partitions the sequence of transmissions $\mathbf{x}$ into subsequences of length $i$, so that

$$\mathbf{x} = [\mathbf{x}^{(1)}, \mathbf{x}^{(2)}, \ldots], \tag{48}$$

where

$$\mathbf{x}^{(j)} = [x_{i(j-1)+1}, x_{i(j-1)+2}, \ldots, x_{ij}]. \tag{49}$$

(For convenience, we will assume that the length of $\mathbf{x}$ is a multiple of $i$.) Note that this partition is not the same as the one in (35). These sequences are then assigned to *independent channels*, and the receiver is made aware of which molecule passed through which channel. More formally, recalling from the notation in Section II-C that $\mathbf{u}$ is the vector formed by arranging the receptions in order of departure time, i.e., $u_i$ is the reception corresponding to $x_i$. Let $\mathbf{u}^{(j)} = [u_{i(j-1)+1}, u_{i(j-1)+2}, \ldots, u_{ij}]$, corresponding to the transmissions in $\mathbf{x}^{(j)}$. Then in this channel, the vector of receptions $\mathbf{y}^*[i]$ is formed by

$$\mathbf{y}^*[i] = [\mathbf{y}^{(1)}, \mathbf{y}^{(2)}, \ldots], \tag{50}$$

where

$$\mathbf{y}^{(j)} = \mathrm{sort}_\tau(\mathbf{u}^{(j)}). \tag{51}$$

Note that we can recover the vector of receptions $\mathbf{y}$ from the original (non-partitioned) channel by sorting $\mathbf{y}^*[i]$, since

$$\mathbf{y} = \mathrm{sort}_\tau(\mathbf{u}) = \mathrm{sort}_\tau(\mathbf{y}^*[i]). \tag{52}$$

Given this partitioned channel, we have the following result:

*Theorem 5:*    1) If $\mathbf{x}$ contains $n$ partitions of length $i$, then

$$f_{\mathbf{Y}^*[i]|\mathbf{X}}(\mathbf{y}^*[i]|\mathbf{x}) = \prod_{j=1}^{n} f_{\mathbf{Y}^{(j)}|\mathbf{X}^{(j)}}(\mathbf{y}^{(j)} \mid \mathbf{x}^{(j)}); \tag{53}$$

and

2) $I(X; Y^*[i]) \geq I(X; Y)$ for all integers $i$.

*Proof:*

1) Under the WI assumptions, molecules propagating through independent channels are equivalent to distinct molecules propagating through the same channel, so this statement follows from Theorem 2.

2) This follows from (52) and the data processing inequality.

$\blacksquare$

From part 1 of Theorem 5, so long as $i$ is small, it is possible to tractably calculate $f_{\mathbf{Y}^*[i]|\mathbf{X}}(\mathbf{y}^*[i]|\mathbf{x})$, and thus it is possible to tractably calculate $I(X; Y[i])$ for small values of $i$ (e.g., using Monte Carlo methods). From part 2 of the theorem, we have a *sequence of upper bounds* on $I(X; Y)$, increasing in complexity as $i$ increases, and approaching the true value of $I(X; Y)$ as $i \to \infty$. Furthermore, we can show the following as a corollary to the theorem:

*Corollary (to Theorem 5):* If $i$ and $j$ are integers, then $I(X; Y^*[ij]) \leq I(X; Y^*[i])$ and $I(X; Y^*[ij]) \leq I(X; Y^*[j])$.

This also follows from the data processing inequality, as groups of $i$ (or $j$) independent channels can be grouped into any integer multiple of $i$ (or, respectively, $j$), and sorted into a block of $ij$ arrivals. Thus, the sequence of upper bounds forms a partial order for general input distributions. We conjecture that this sequence of bounds is monotonic (i.e., that $I(X; Y^*[i]) \leq I(X; Y^*[j])$ for all $i > j$), but we leave the proof of this statement to future work.

## VI. RESULTS AND DISCUSSION

### A. Preliminaries

In Theorem 3, we saw that unconstrained input distributions lead to infinite capacity. To generate more meaningful results, we constrain the input distribution to be *discrete-time* and *binary*. In particular, we quantize time to intervals of length $T$: at the beginning of each interval, we release a single molecule with probability $p_x$, or no molecule with probability $(1 - p_x)$, where the probability of transmission at each interval is statistically independent. (Using our notation from Section IV, one might also say that $\Pr(\eta(\mathbf{x}_j) = 1) = p_x$, and $\Pr(\eta(\mathbf{x}_j) = 0) = 1 - p_x$, for all $j$.) This is a practical constraint, and is analogous to a peak power constraint in a conventional communication system, as the transmitter cannot release more than one molecule every $T$ seconds. We make no claim that this form of input distribution is optimal, and we leave the difficult problem of optimizing the input distribution to future work.

We use the WI model to obtain all our experimental results. From (25), this model has two parameters: $d$ and $\sigma^2$, and we assign $d = \sigma^2 = 1$ (however, since these two parameters always appear as $d^2/\sigma^2$, any system with $d^2/\sigma^2 = 1$ would have the same performance). We pick these numbers only to illustrate the relative performances of the bounds.

Using our approximate models from Section IV, the counting intervals are the same as the intervals between transmissions. Meanwhile, we use the following method to select the Poisson "noise" parameter $\lambda$. The probability that a molecule arrives in the $i$ observation intervals of length $T$ is given by

$$p_a = \sum_{j=0}^{i-1} p_{a,j}(0), \tag{54}$$

where $p_{a,i}(0)$ is given by (40). Thus, the probability that the molecule is "lost" is given by $(1 - p_a)$. The expected number of "lost" molecules generated per interval is then $p_x(1 - p_a)$. In the steady state, this must be the same as the expected number of "lost" molecules arriving per interval. Thus, we set

$$\lambda = p_x(1 - p_a) = p_x \left( 1 - \sum_{j=0}^{i-1} p_{a,j}(0) \right). \tag{55}$$

Since $p_a \to 1$ (and thus $\lambda \to 0$) as $i \to \infty$, this setting of $\lambda$ is consistent with the convergence of the sequence of lower bounds to the true mutual information of the counting detector.

### B. Results

We begin with results illustrating the upper and lower bounds together, which are given in Figures 3 and 4, showing mutual information in bits per unit time, and bits per molecule, respectively, with respect to $p_x$. The "unit time" is the interval length $T$, which is set to $T = 2.198$, the amount of time (given $d^2/\sigma^2 = 1$) such that the probability of arrival in the first interval is $p_{a,0}(0) = 0.5$ We depict the first four lower bounds and the first two upper bounds. We see a significant improvement of performance in the bounds moving from the first to the second bound, and all the bounds monotonically improve with order. Furthermore, practical information rates (especially in terms of bits per molecule) are clearly possible. As expected from Theorem 3, low values of $p_x$ lead to high information rates per molecule.

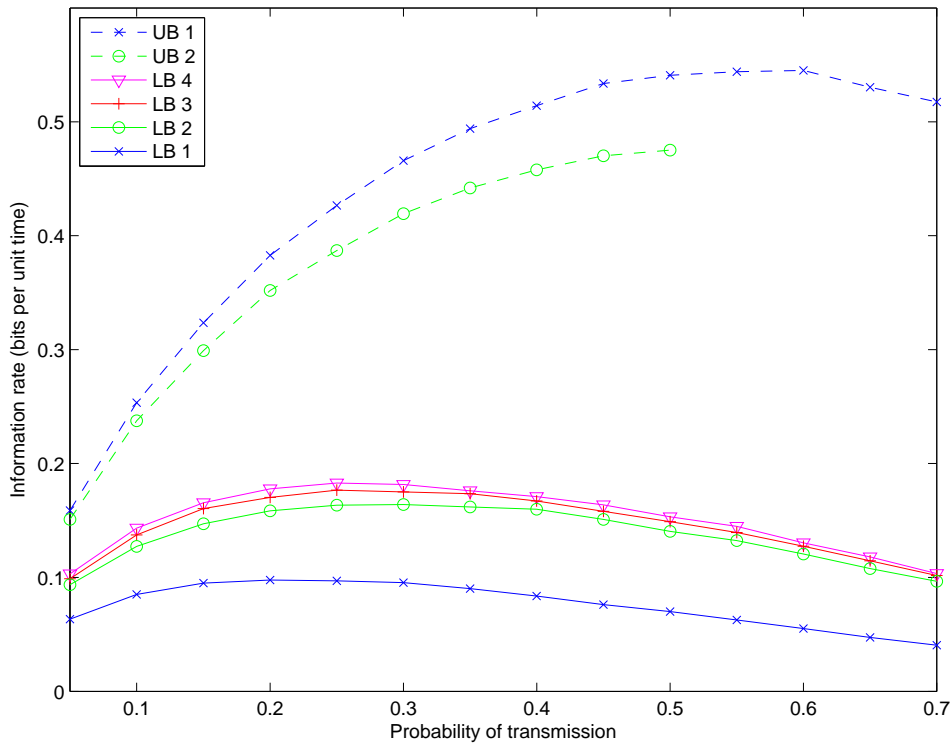

Fig. 3.  Upper bounds (UB) and lower bounds (LB) on mutual information per unit time with respect to $p_x$.

Since the lower bounds are known to be achievable, whereas the upper bounds are not, we focus additional attention on the performance of the lower bounds. In Figures 5 and 6, we plot the information rate of the lower bound per unit time where the interval lengths are $T = 1.068$ and $T = 5.390$, respectively; for which the probabilities of arrival in the first interval are $p_{a,0}(0) = 0.333$ and $p_{a,0}(0) = 0.667$, respectively. (However, for fair comparison, the curves are still plotted as bits per unit of time $T = 2.198$.) In Figures 7 and 8, we also give results for the same two systems in terms of bits per molecule. From Figures 5 and 7, we see a significant improvement in performance from each additional order of the bound, which is intuitive since the arrival probability in the first interval is small; conversely, little performance improvement is observed beyond the second order in Figures 6 and 8.

## C. Discussion

Our results are primarily intended to illustrate the feasibility of our upper and lower bounds in estimating the true mutual information in the channel. To that end, from Figures 3-4, we see that our upper and lower bounds are reasonably successful in giving an idea of the true mutual information in molecular communication channels; in the worst case, the upper and lower bounds agree at least in order of magnitude. Nonetheless, more work needs to be done to narrow the gap between the upper and lower bounds. Furthermore, it is interesting that the maximizing value of $p_x$ is considerably different in the case of the upper and lower bounds. Meanwhile, our results from Figures 5-8 indicate that the achievable lower bound produces useful results which obey our intuition.

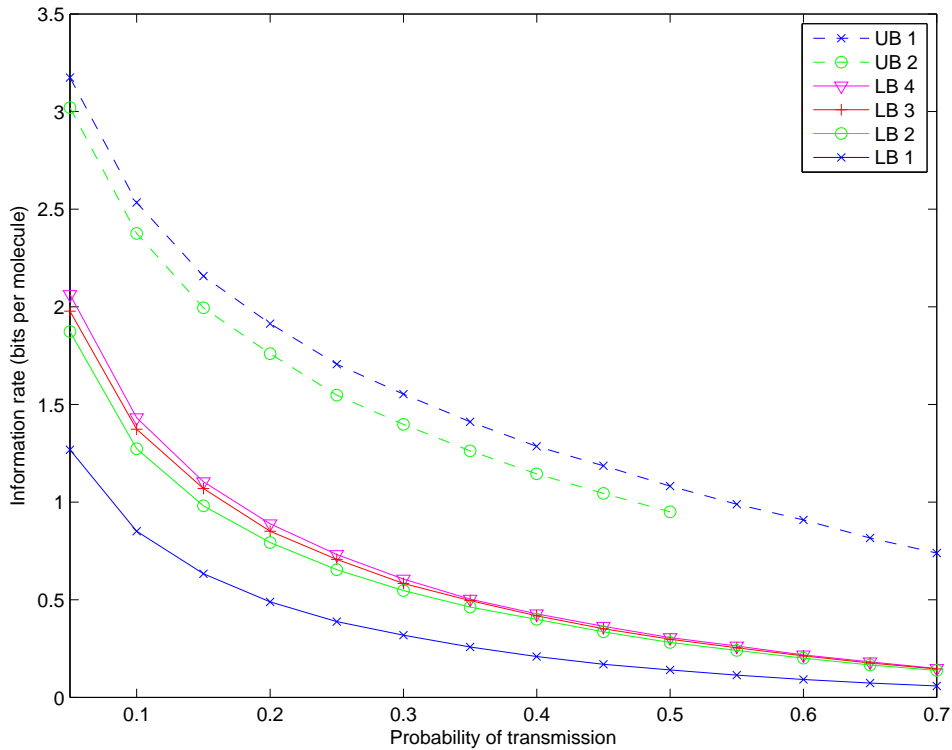

Fig. 4. Upper bounds (UB) and lower bounds (LB) on mutual information per molecule with respect to $p_x$.

It is natural to ask whether we expect the true mutual information to lie closer to the upper or lower bound. For the lower bound, we see that the gap between successive curves is smaller as the order increases; thus, it is reasonable to believe that the true mutual information for the counting detector is not much greater than the highest-order bounds we have provided. However, as we showed in Example 1, the counting detector's mutual information is likely smaller than that of the ideal detector. For the ideal detector, the upper bounds (which tend to require high computational complexity to calculate) show a significant gap from first order to second order; thus, we cannot surmise whether the true mutual information is close to the upper bound.

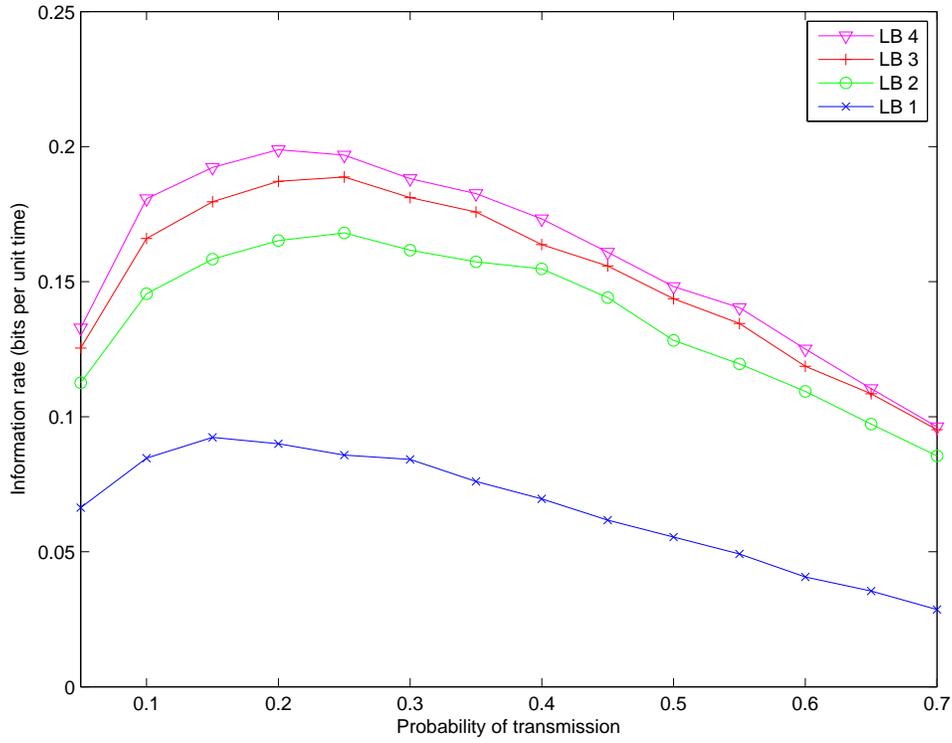

Fig. 5. Lower bounds (LB) on mutual information per unit time with respect to $p_x$, for $T = 1.068$.

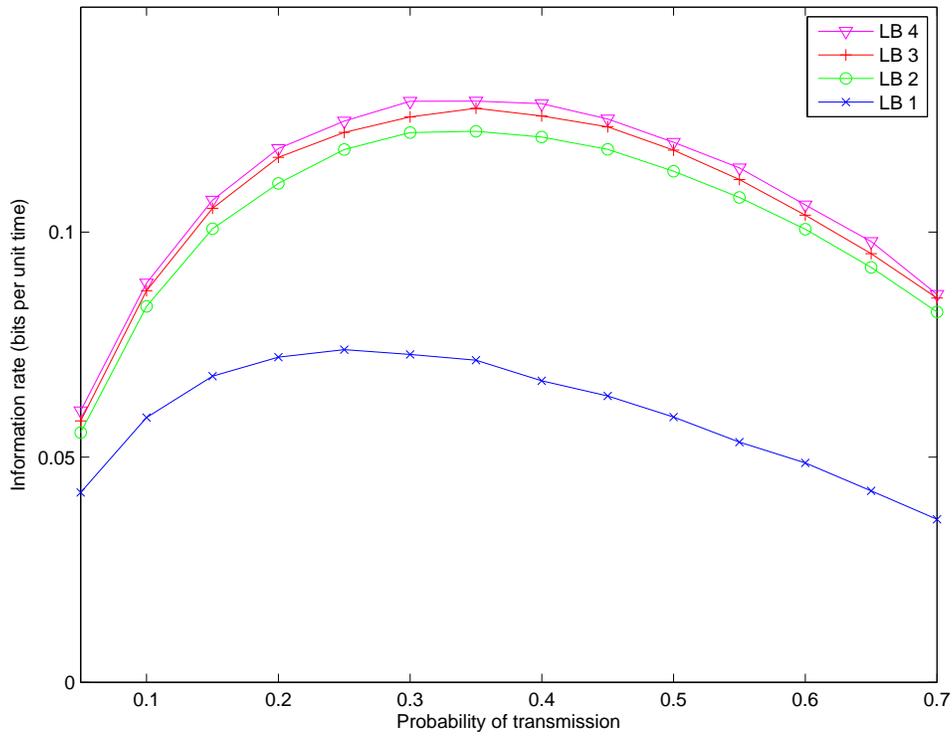

Fig. 6. Lower bounds (LB) on mutual information per unit time with respect to $p_x$, for $T = 5.390$.

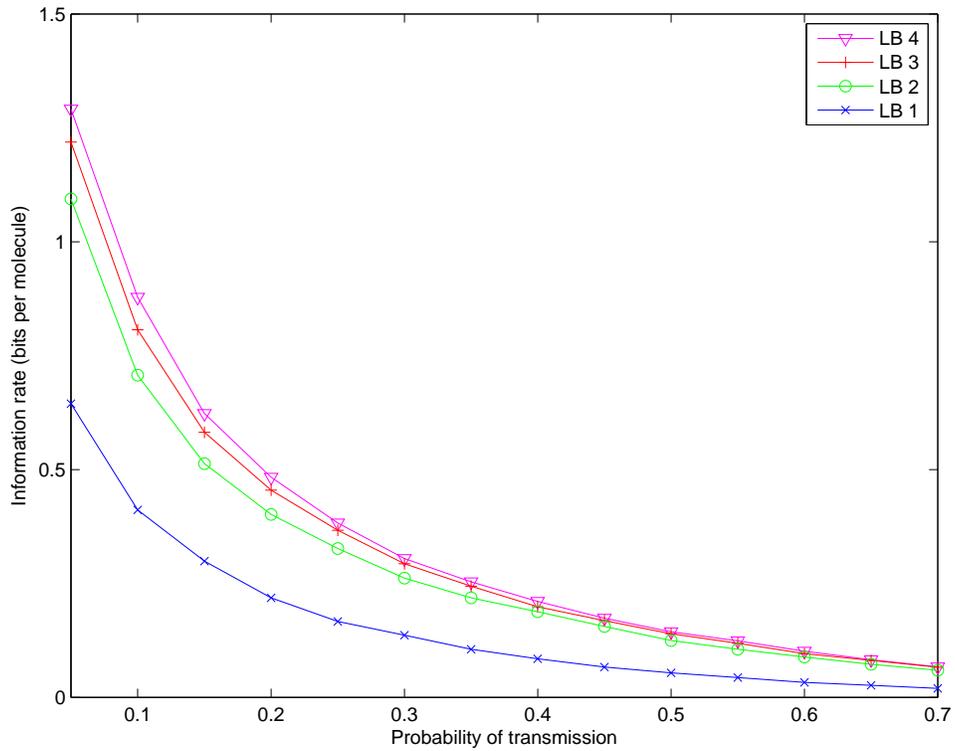

Fig. 7. Lower bounds (LB) on mutual information per molecule with respect to $p_x$, for $T = 1.068$.

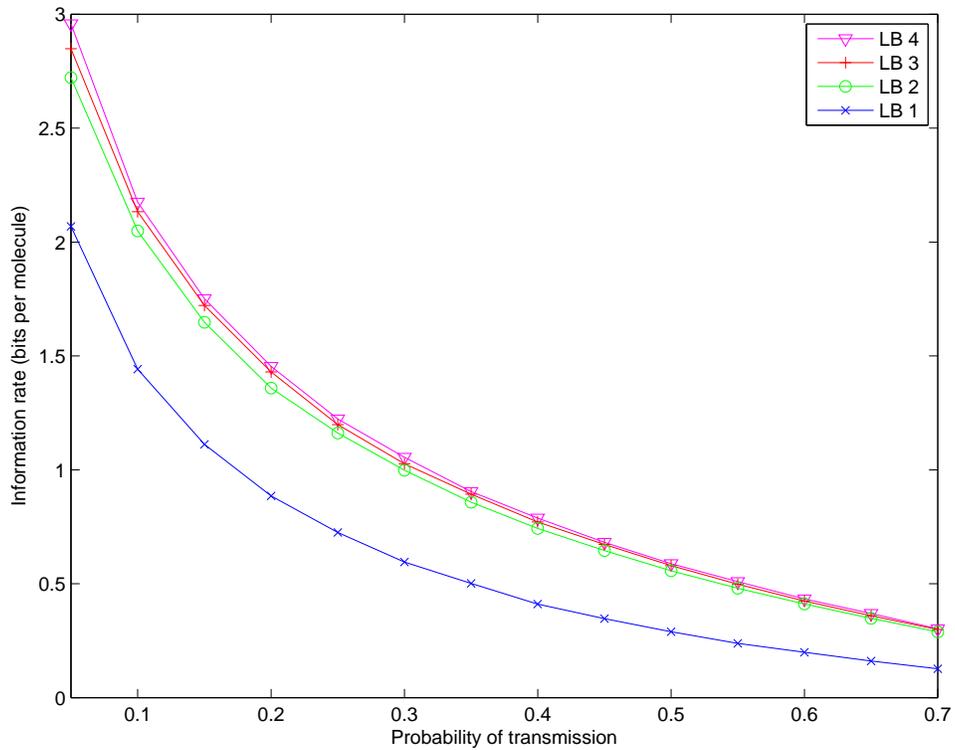

Fig. 8. Lower bounds (LB) on mutual information per molecule with respect to $p_x$, for $T = 5.390$.

## VII. Conclusion

In this paper, we have provided an information-theoretic and communication-theoretic basis for molecular communication. There exists a vast body of tools and techniques in dealing with novel communication channels, and our modeling results provide an avenue between molecular communication and traditional communication theory. Furthermore, we have provided bounds on information rate for molecular communication, which are feasible for estimating the mutual information of such channels. Our work opens up a vast array of new problems in coding and information theory for molecular communication, such as the problems of finding improved bounds, optimal input distributions, and models for related systems.

## Appendix

### A. Derivation of (16)

From (14), we have the CDF form of the Bapat-Beg theorem. The PDF form is derived as follows. We know that

$$f_{\mathbf{Z}}(\mathbf{z}) = \frac{\partial}{\partial z_1} \frac{\partial}{\partial z_2} \cdots \frac{\partial}{\partial z_n} \Pr(Z_1 < z_1, Z_2 < z_2, \ldots, Z_n < z_n), \tag{56}$$

so from (14)-(15), and since $\mathrm{per}(\mathbf{M}) = \mathrm{per}(\mathbf{M}^T)$, we can write

$$\begin{aligned} f_{\mathbf{Z}}(\mathbf{z}) &= \frac{\partial}{\partial z_1} \frac{\partial}{\partial z_2} \cdots \frac{\partial}{\partial z_n} \sum_{\pi \in \mathcal{P}_n} F_{W_{\pi(1)}}(z_1) \prod_{i=2}^{n} (F_{W_{\pi(i)}}(z_i) - F_{W_{\pi(i)}}(z_{i-1})) \\ &= \sum_{\pi \in \mathcal{P}_n} \frac{\partial}{\partial z_1} \frac{\partial}{\partial z_2} \cdots \frac{\partial}{\partial z_n} F_{W_{\pi(1)}}(z_1) \prod_{i=2}^{n} (F_{W_{\pi(i)}}(z_i) - F_{W_{\pi(i)}}(z_{i-1})). \end{aligned} \tag{57}$$

Considering each term in the permanental sum from (57), we can show by induction that

$$\frac{\partial}{\partial z_1} \frac{\partial}{\partial z_2} \cdots \frac{\partial}{\partial z_n} F_{W_{\pi(1)}}(z_1) \prod_{i=2}^{n} (F_{W_{\pi(i)}}(z_i) - F_{W_{\pi(i)}}(z_{i-1})) = \prod_{i=1}^{n} f_{W_{\pi(i)}}(z_i). \tag{58}$$

To do so, note first that $\frac{\partial}{\partial z_1} F_{W_{\pi(1)}}(z_1) = f_{W_{\pi(1)}}(z_1)$, by definition. Now, in the inductive step, if

$$\frac{\partial}{\partial z_1} \frac{\partial}{\partial z_2} \cdots \frac{\partial}{\partial z_{n-1}} F_{W_{\pi(1)}}(z_1) \prod_{i=2}^{n-1} (F_{W_{\pi(i)}}(z_i) - F_{W_{\pi(i)}}(z_{i-1})) = \prod_{i=1}^{n-1} f_{W_{\pi(i)}}(z_i), \tag{59}$$

then

$$\frac{\partial}{\partial z_1}\frac{\partial}{\partial z_2}\cdots\frac{\partial}{\partial z_{n-1}}\frac{\partial}{\partial z_n}F_{W_{\pi(1)}}(z_i)\prod_{i=2}^{n}(F_{W_{\pi(i)}}(z_i)-F_{W_{\pi(i)}}(z_{i-1})) \tag{60}$$

$$=\frac{\partial}{\partial z_1}\frac{\partial}{\partial z_2}\cdots\frac{\partial}{\partial z_{n-1}}\frac{\partial}{\partial z_n}F_{W_{\pi(1)}}(z_i)\prod_{i=2}^{n-1}(F_{W_{\pi(i)}}(z_i)-F_{W_{\pi(i)}}(z_{i-1}))F_{W_{\pi(n)}}(z_n)$$

$$-F_{W_{\pi(1)}}(z_i)\prod_{i=2}^{n-1}(F_{W_{\pi(i)}}(z_i)-F_{W_{\pi(i)}}(z_{i-1}))F_{W_{\pi(n)}}(z_{n-1}) \tag{61}$$

$$=\left[\frac{\partial}{\partial z_1}\frac{\partial}{\partial z_2}\cdots\frac{\partial}{\partial z_{n-1}}F_{W_{\pi(1)}}(z_i)\prod_{i=2}^{n-1}(F_{W_{\pi(i)}}(z_i)-F_{W_{\pi(i)}}(z_{i-1}))\right]\frac{\partial}{\partial z_n}F_{W_{\pi(n)}}(z_n) \tag{62}$$

$$=\left[\prod_{i=1}^{n-1}f_{W_{\pi(i)}}(z_i)\right]\frac{\partial}{\partial z_n}F_{W_{\pi(n)}}(z_n) \tag{63}$$

$$=\prod_{i=1}^{n}f_{W_{\pi(i)}}(z_i), \tag{64}$$

where (62) follows from the fact that the second term in (61) is independent of $z_n$, (63) follows from the inductive hypothesis in (59), and (64) follows from the definition of the PDF. Thus, from (64), we have that

$$f_{\mathbf{Z}}(\mathbf{z})=\sum_{\pi\in\mathcal{P}_n}\prod_{i=1}^{n}f_{W_{\pi(i)}}(z_i), \tag{65}$$

which is the permanent of a matrix whose $(i,j)$th entry is $f_{W_i}(z_j)$.

Finally, (16) follows by substituting $F_{W_i}(z_j)$ with $F_{Y_i|X_j}(y_i|x_j)$. Although $y_i=(m_i,s_i,\tau_i')$, and the order statistics $\mathbf{y}$ are sorted with respect to $\tau_i$, the Bapat-Beg theorem admits order statistics where the ordering is with respect to a single component of a vector random variable, such as $y_i$. Finally, the result is obtained by differentiating with respect to $y_1, y_2, \ldots, y_n$ rather than $z_1, z_2, \ldots, z_n$.

## REFERENCES

[1] S. P. Brown and R. A. Johnstone, "Cooperation in the dark: Signalling and collective action in quorum-sensing bacteria," *Proceedings of the Royal Society of London B*, vol. 268, pp. 961–965, 2001.

[2] S. Hiyama, Y. Moritani, T. Suda, R. Egashira, A. Enomoto, M. Moore, and T. Nakano, "Molecular communication," in *Proc. 2005 NSTI Nanotechnology Conference*, pp. 391–394, 2005.

[3] H. Kitano, "Systems biology: A brief overview," *Science*, vol. 295, pp. 1662–1664, 1 Mar. 2002.

[4] R. Weiss, S. Basu, S. Hooshangi, A. Kalmbach, D. Karig, R. Mehreja, and I. Netravali, "Genetic circuit building blocks for cellular computations, communications, and signal processing," *Natural Computing*, vol. 2, pp. 47–84, Mar. 2003.

[5] R. Weiss and T. Knight, "Engineered communications for microbial robotics," in *Proc. 6th International Meeting on DNA Based Computers*, 2000.

[6] T. Nakano, T. Suda, M. Moore, R. Egashira, A. Enomoto, and K. Arima, "Molecular communication for nanomachines using intercellular calcium signalling," in *Proc. 5th IEEE Conference on Nanotechnology*, pp. 478–481, 2005.

[7] T. Nakano, T. Suda, T. Kojuin, T. Haraguchi, and Y. Hiraoka, "Molecular communication through gap junction channels: System design, experiments and modeling," in *Proc. 2nd International Conference on Bio-Inspired Models of Network, Information, and Computing Systems,* Budapest, Hungary, 2007.

[8] A. Enomoto, M. Moore, T. Nakano, R. Egashira, T. Suda, A. Kayasuga, H. Kojima, H. Sakibara, and K. Oiwa, "A molecular communication system using a network of cytoskeletal filaments," in *Proc. 2006 NSTI Nanotechnology Conference*, pp. 725–728, 2006.

[9] S. Hiyama, Y. Moritani, and T. Suda, "A biochemically engineered molecular communication system," in *Proc. 3rd International Conference on Nano-Networks,* Boston, MA, USA, 2008.

[10] A. Cavalcanti, T. Hogg, B. Shirinzadeh, and H. C. Liaw, "Nanorobot communication techniques: A comprehensive tutorial," in *IEEE Intl. Conf. on Control, Automation, Robotics and Vision,* Singapore, 2006.

[11] T. Berger, "Living information theory (Shannon lecture)," in *Proc. IEEE International Symposium on Information Theory,* Lausanne, Switzerland, 2002.

[12] D. Blackwell, *Information theory*, pp. 182–193. Modern mathematics for the engineer: Second series, New York: McGraw-Hill, 1961.

[13] H. Permuter, P. Cuff, B. V. Roy, and T. Weissman, "Capacity and zero-error capacity of the chemical channel with feedback," in *Proc. IEEE International Symposium on Information Theory,* Nice, France, 2007.

[14] V. Anantharam and S. Verdú, "Bits through queues," *IEEE Trans. Inform. Theory*, vol. 42, pp. 4–18, Jan. 1996.

[15] R. Sundaresan and S. Verdú, "Capacity of queues via point-process channels," *IEEE Trans. Inform. Theory*, vol. 52, pp. 2697–2709, Jun. 2006.

[16] D. M. Arnold, H.-A. Loeliger, P. O. Vontobel, A. Kavčić, and W. Zeng, "Simulation-based computation of information rates for channels with memory," *IEEE Trans. Inform. Theory*, vol. 52, pp. 3498–3508, Aug. 2006.

[17] I. Karatzas and S. E. Shreve, *Brownian Motion and Stochastic Calculus (2nd edition)*. New York: Springer, 1991.

[18] R. B. Bapat and M. I. Beg, "Order statistics for nonidentically distributed variables and permanents," *Sankhya (Ser. A)*, vol. 51, no. 1, pp. 79–93, 1989.

[19] H. Minc, *Permanents*, vol. 6 of *Encyclopedia of Mathematics and its Applications*. Reading, MA: Addison-Wesley, 1978.

[20] L. G. Valiant, "The complexity of computing the permanent," *Theoretical Computer Science*, vol. 8, pp. 189–201, 1979.

[21] H. J. Ryser, *Combinatorial Mathematics*. The Carus Mathematical Monographs no. 14, Mathematical Association of America, 1963.

[22] M. Jerrum, A. Sinclair, and E. Vigoda, "A polynomial-time approximation algorithm for the permanent of a matrix with nonnegative entries," *Journal of the ACM*, vol. 51, pp. 671–697, Jul. 2004.

[23] K. H. Rosen and J. G. Michaels, *Handbook of discrete and combinatorial mathematics*. CRC Press, 2000.

[24] S. Goldstein, "Mechanical models of Brownian motion," *Lecture Notes in Physics*, vol. 153, pp. 21–24, 1982.

[25] P. Harremöes, "Convergence to Poisson distribution," in *Proc. Conference in Computer Science and Information Technologies,* Yerevan, Armenia, pp. 224–228, 2001.